\documentclass[useAMS,usenatbib,letters]{mn2e}
\usepackage{journal_abbreviations}
\usepackage{graphicx, amsmath, amssymb}
\usepackage{enumerate}
\usepackage{float}

\usepackage{mathrsfs}
\usepackage{graphics}

\title[The structural evolution of massive galaxies]{The average structural evolution of massive galaxies can be reliably estimated using cumulative galaxy number densities}
\author[Clauwens et al.]{Bart Clauwens$^{1,2}$\thanks{E-mail: clauwens@strw.leidenuniv.nl}, Allison Hill$^{1}$, Marijn Franx$^{1}$, Joop Schaye$^{1}$\\
$^{1}$Leiden Observatory, Leiden University, PO Box 9513, 2300 RA Leiden, The Netherlands\\$^{2}$Instituut-Lorentz for Theoretical Physics, Leiden University, 2333 CA Leiden, The Netherlands\\}

\begin{document}

\date{Accepted . Received 2017 April 11; in original form 2017 March 6.}

\pagerange{\pageref{firstpage}--\pageref{lastpage}}

\maketitle

\label{firstpage}

\begin{abstract}
Galaxy evolution can be studied observationally by linking progenitor and descendant galaxies through an evolving cumulative number density selection. This procedure can reproduce the expected evolution of the median stellar mass from abundance matching. However, models predict an increasing scatter in main progenitor masses at higher redshifts, which makes galaxy selection at the median mass unrepresentative. Consequently, there is no guarantee that the evolution of other galaxy properties deduced from this selection are reliable. Despite this concern, we show that this procedure approximately reproduces the evolution of the average stellar density profile of main progenitors of $M\approx10^{11.5}{\rm M_{\odot}}$ galaxies, when applied to the EAGLE hydrodynamical simulation. At $z\gtrsim3.5$ the aperture masses disagree by about a factor two, but this discrepancy disappears when we include the expected scatter in cumulative number densities. The evolution of the average density profile in EAGLE broadly agrees with observations from UltraVISTA and CANDELS, suggesting an inside-out growth history for these massive galaxies over $0\lesssim z \lesssim 5$. However, for $z\lesssim 2$ the inside-out growth trend is stronger in EAGLE. We conclude that cumulative number density matching gives reasonably accurate results when applied to the evolution of the mean density profile of massive galaxies.
\end{abstract}


\section{Introduction}
\label{SectionIntroduction}

An interesting challenge for the study of galaxy evolution is to find a method to distill the evolution of a typical galaxy from that of the total observed galaxy population. One way to do this, is to rank the observed galaxies at each redshift according to their stellar mass and then assign a unique cumulative number density (hereafter CND) to each galaxy, defined as the comoving number density ($\rm{Mpc^{-3}}$) of galaxies at that redshift that are more massive than the given galaxy. The simplest assumption is then that galaxies evolve along a constant CND. This method has been used to study, among other things, the evolution of stellar masses, star formation rates and stellar density profiles of Milky Way-like and massive galaxies \citep{Loeb03,Dokkum10,Papovich11,Dokkum13,Patel13,Lundgren14,Morishita15}.

However, the assumption of a constant CND is a very crude approximation to the expected evolution in a $\Lambda$CDM cosmology, which is inherently stochastic in nature and cannot conserve galaxy numbers due to merging.
\citeauthor{Behroozi13} (\citeyear{Behroozi13}, hereafter B13) give a more accurate prediction for the expected evolution of the CND, which accounts for mergers. This prediction has been applied to observations by \citet{Marchesini14} and  \citet{Vulcani16} to study, among other properties, the stellar mass, star formation rate and environments of the main progenitors of massive galaxies. \citet{Papovich15} use a similar method to study the evolution of the stellar density profiles of main progenitors of both massive and Milky Way-mass galaxies.

B13 base their prediction for the evolution of the median CND on abundance matching of observed galaxies to halos in the Bolshoi dark matter simulation. The median is taken at each redshift and represents the typical main progenitor. They find a significant increase of the median CND of $\approx0.16$ dex per $\Delta z$ when tracing main progenitors\footnote{B13 have made their exact results available at http://code.google.com/p/nd-redshift/.} and a much smaller redshift dependence when tracing descendants\footnote{One should keep in mind that, for descendants, the median CND is defined only with respect to the surviving galaxies and might evolve more steeply when the merged galaxies are included in the median.}. They also find a large scatter ($\approx 0.7$ dex) around this median CND that increases with $\Delta z$.

The expected evolution of the median number density of the progenitors/descendants can be accounted for when applied to observations \citep{Marchesini14,Papovich15,Vulcani16}. The large scatter is more problematic, because it potentially defeats the purpose of the method to reliably identify progenitors/descendants. It would be more accurate to also sample the scatter in the CND. However, when sampling such a wide distribution of galaxy masses, there are many galaxies to choose from and thus also many ways in which to choose galaxies with the wrong properties. This would not matter if there were no additional independent correlations between the properties of descendants and progenitors. It is however possible that, for example, progenitors of similar stellar mass with more/less centrally concentrated density profiles tend to form more/less massive descendants. The assumption that the evolution of galaxy properties can be reliably estimated from a stellar mass-selected galaxy sample, might hold to a differing degree for different galaxy properties.

For this reason it is important to test the CND matching technique on semi-analytic models and hydrodynamical simulations of galaxy formation in a context where these additional independent correlations might appear. In a previous paper \citep{Clauwens16} we reported such a correlation between the $z=0$ star formation rate of Milky way-mass galaxies and the stellar mass of their main progenitors in the EAGLE hydrodynamical simulation \citep{Schaye15,Crain15}: main progenitors of passive galaxies tend to be a factor 2.5 more massive at $z=2$ than main progenitors of active galaxies with the same stellar mass.

In this paper we investigate the extent to which CND matching can retrieve the stellar density profiles of the main progenitors of massive galaxies out to $z=5$ in the EAGLE simulation. We base our galaxy selection on the observational work of \citet{Hill16} and we conclude with a direct comparison of simulation and observations.

Other works that analyse the performance of CND matching in semi-analytic models and hydrodynamic simulations include \citet{Leja13,Torrey15b,Torrey16, Mundy15,Henriques15,Terrazas16,Jaacks16,Voort16, Wellons16}. One advantage of using the EAGLE simulation for our analysis, is that it matches the observed evolution of the galaxy stellar mass function (GSMF) quite well \citep{Furlong15}. Therefore, the B13 prescription, which is based on the observed GSMF through abundance matching, is expected to work for the EAGLE simulation, at least in retrieving the median stellar mass evolution.

\section{Method}
\label{SectionMethod}

For this work we use the $\rm (100 \, Mpc)^3$ sized reference run of the EAGLE hydrodynamical simulation: Ref-L100N1504. All the used data is publicly available \citep{McAlpine16}. The simulation includes radiative cooling and heating, and stochastic feedback from active galactic nuclei and stars (the latter depending on the local density and metallicity). The simulation has been calibrated to reproduce the current GSMF and galaxy sizes. The effective resolution is set by an initial gas particle mass of $1.6 \times 10^6 \rm{M_{\odot}}$, a maximal gravitational force softening of 700 pc and an effective temperature pressure floor of 8000 K for the interstellar medium \citep{Schaye08}. This means that  by design the simulation does not give cold thin disks. The minimal disk height is roughly 1 proper kpc (pkpc). In some cases we will give results on a 1 pkpc scale, but keep in mind that this is stretching the domain of applicability of the simulation. Certainly these results will be much less robust then those on scales of 3 pkpc and larger.

We base the galaxy selection in EAGLE on the observational work of \citet{Hill16}, who use stacked observations from UltraVISTA and CANDELS out to $z=5$ to study the structural evolution of the progenitors of massive galaxies. Their descendants are selected at $0.2<z<0.5$ to have a stellar mass of $10^{11.5} {\rm M_{\odot}}$. Main progenitor stacks are made in redshift intervals out to $z=5$ along the evolving CND prescribed by B13.

In EAGLE we select all the galaxies at $z=0.37$ that have a stellar mass of $11.4<{\rm Log}(M/\rm{M_{\odot}})<11.6$ within a 70 pkpc 3D aperture, comparable with the 75 pkpc 3D aperture aimed for in the observations. This gives a sample of 24 descendant galaxies. Throughout this paper we will use the 3D stellar aperture masses within 1, 3, 5, 20 and 70 pkpc as a coarse sampling of the stellar density profiles.

We follow the main progenitor evolution of the aperture masses for both the true main progenitors of these 24 galaxies and for the 24 galaxies closest to the median CND from B13. Lastly, we make a  `fiducial' selection of 24 galaxies for which the expected variance in the main progenitor CNDs is taken into account. For this we fit an evolving lognormal distribution of CNDs, based on the $68^{\rm th}$ percentile range given by B13. We sample the lognormal distribution at regular cumulative probabilities (multiples of 1/25). Recently \citet{Torrey16} and \citet{Wellons16} also advocated the use of a lognormal distribution of CNDs. 

\section{Results}
\label{SectionResults}

\begin{figure*}
\includegraphics[height=0.27\textwidth]{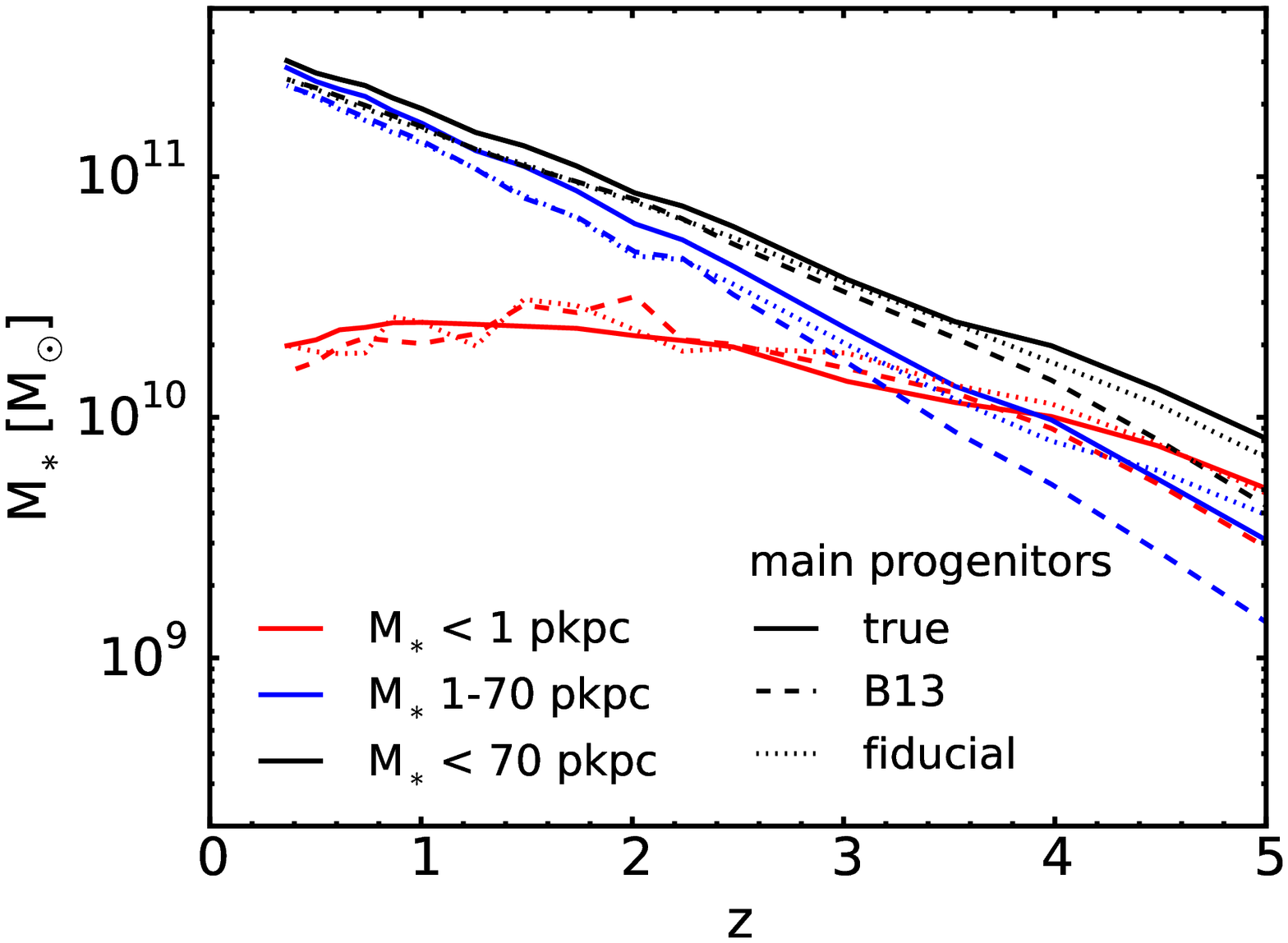}
\includegraphics[height=0.27\textwidth]{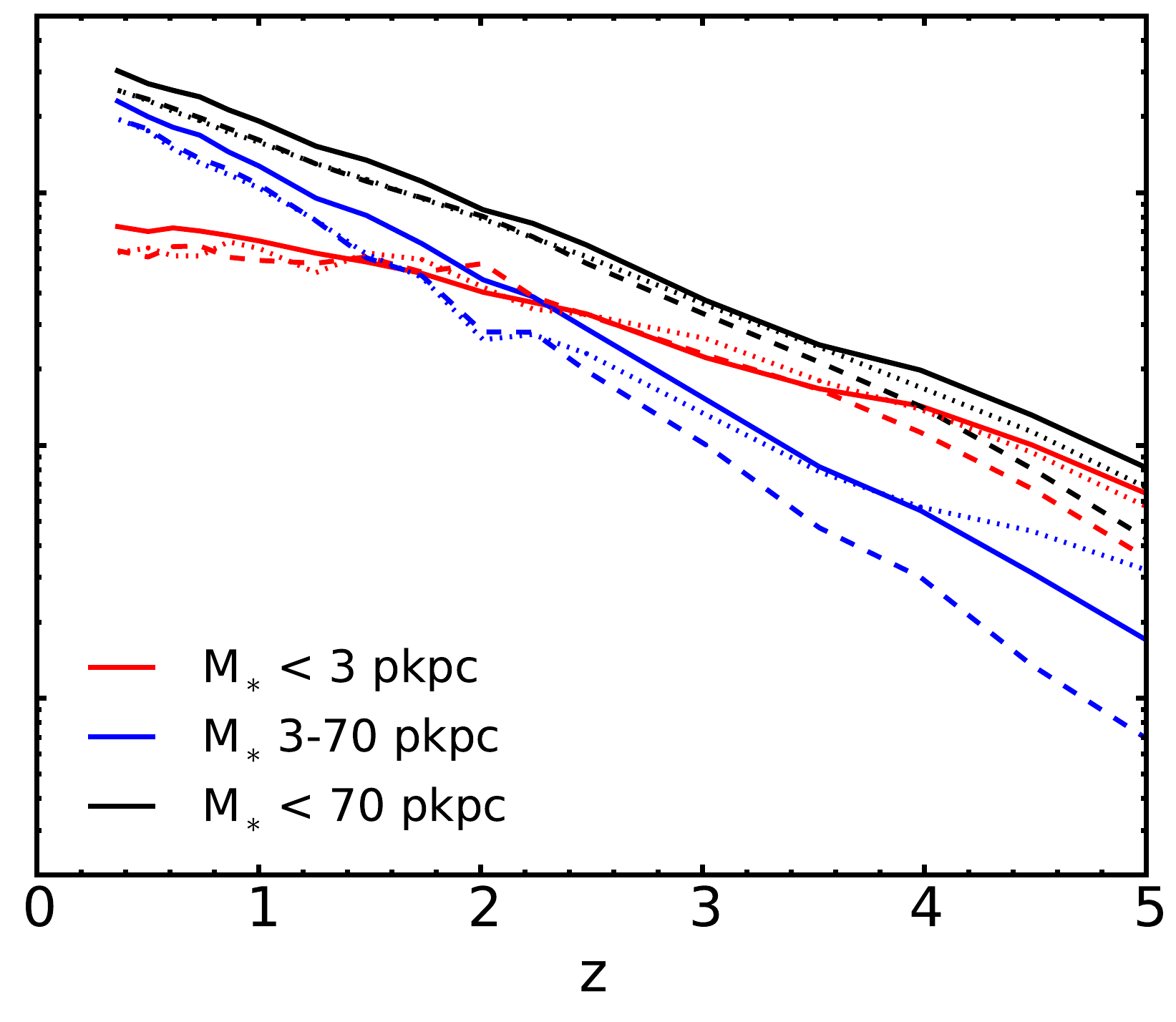}
\includegraphics[height=0.27\textwidth]{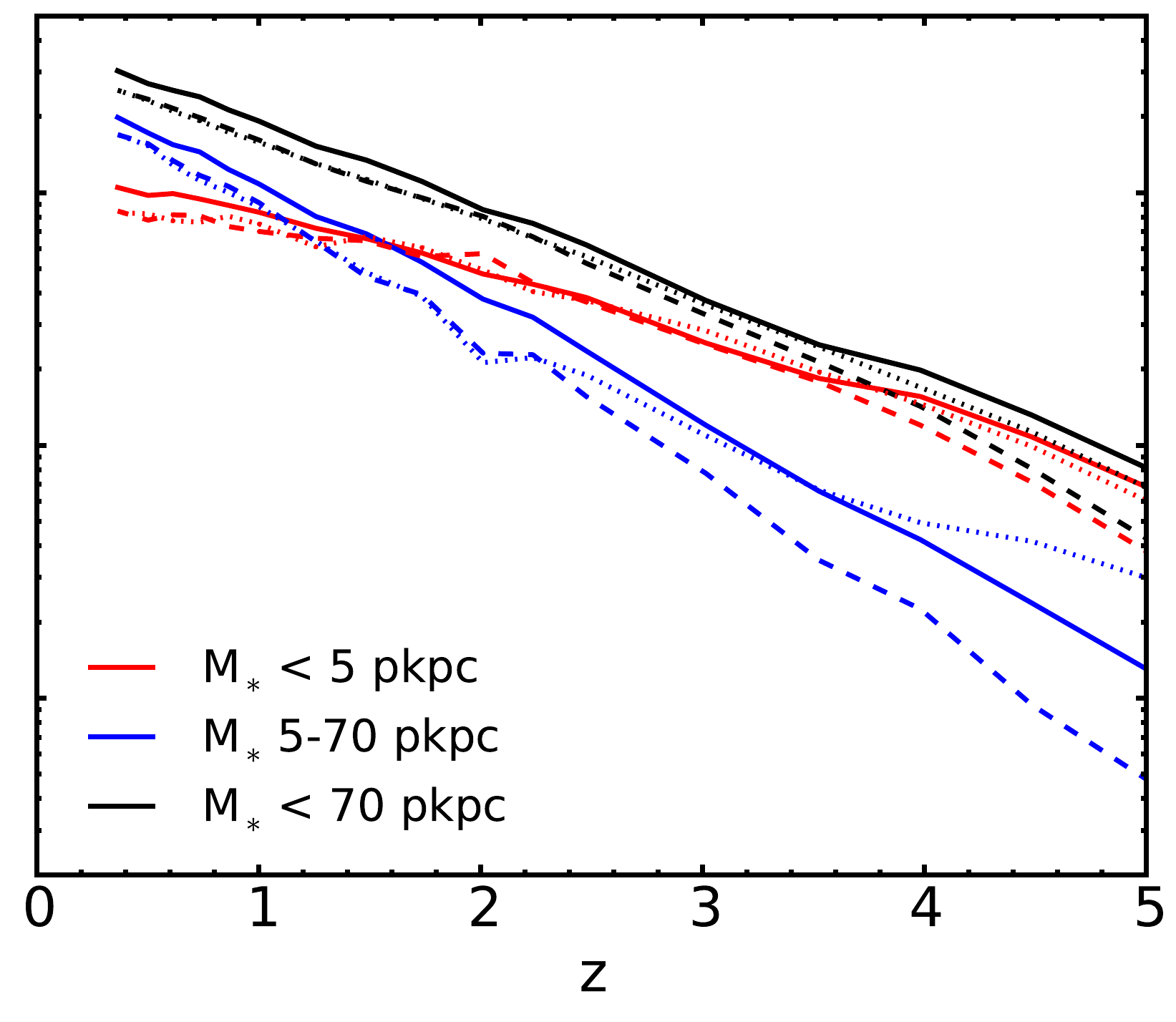}
\caption{The evolving average stellar mass within a fixed 3D proper aperture. The apertures for the red curves in the left, middle and right panels are 1,3 and 5 pkpc, respectively. Black curves denote the aperture masses within 70 pkpc, which are the same in all panels. Blue curves denote the difference between inner and outer apertures (black and red). Different line styles denote different evolving galaxy selections within the EAGLE simulation. The true main progenitor evolution is depicted by solid curves. Dashed curves represent a selection based on the expected B13 median CND, whereas dotted curves represent the `fiducial' selection, which also includes the scatter in the CND evolution from B13. In all panels the `B13' curves agree remarkably well with the `main progenitors', with an increasing discrepancy towards higher redshifts. The agreement for the `fiducial' curves is even better.}
\label{Figure1}
\end{figure*}

Figure \ref{Figure1} compares the evolution of the average aperture masses for the true main progenitors in EAGLE to the evolution obtained with the `B13' selection and the `fiducial' selection. We use average masses rather than median masses because this is equivalent to what one gets from stacking observations. Remarkably, the `B13' aperture mass evolution (dashed curves) agrees well with the main progenitor aperture mass evolution (solid curves) for all apertures. The agreement is especially good below $z\approx3.5$.  Above this redshift the difference between the main progenitor and the `B13'  mass within a 70 pkpc aperture increases to $\approx0.3$ dex. A similar divergence appears for the smaller apertures.

\begin{figure}
\includegraphics[width=\columnwidth]{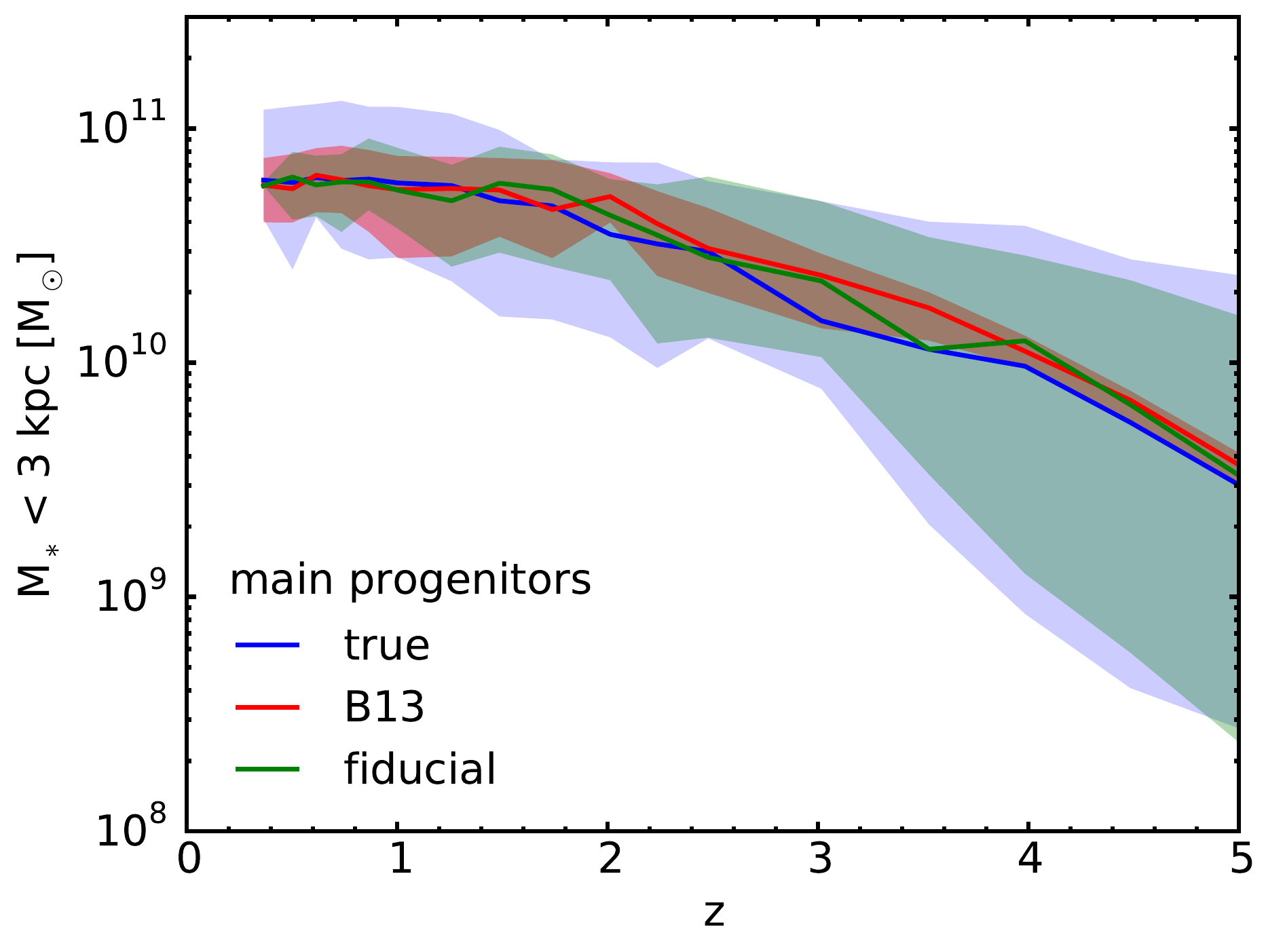}
\caption{The evolving median stellar mass within a 3 pkpc 3D radius. Solid lines show the evolving medians and shaded regions show the 10\%-90\% percentiles. The main progenitor selection is shown in blue, the  `B13' selection in red and the `fiducial' selection is shown in green. Plots for 1 or 5 pkpc apertures look qualitatively similar (not shown) with a more extended 10\%-90\% range for the 1 pkpc aperture. The `fiducial' selection reproduces the true spread in progenitor aperture masses at high redshifts.}
\label{Figure2}
\end{figure}

At $z\gtrsim2.5$ the `fiducial' aperture masses (dotted curves) are more accurate than the `B13' aperture masses (dashed curves). Qualitatively we can understand this because the higher the redshift, the larger the scatter in the true main progenitor masses, and the more important it becomes to model this scatter accurately.

The average density profile of the main progenitors of $M_{*}=10^{11.5} {\rm M_{\odot}}$ galaxies is thus well approximated in EAGLE by using the `B13' method. Assuming that the real Universe resembles the EAGLE simulation in this regard, this means that stacking galaxy images along the B13 CND is a reliable way to estimate the average growth of the density profile of galaxies with this mass. An important question is whether the average density profile is representative of the individual galaxies that compose the stack.

Figure \ref{Figure2} shows the 10\%-90\% range of the main progenitor stellar masses in the 3 pkpc aperture (blue shaded area). The `B13' method (red shaded area), by construction, does not try to match this and indeed does not. The `fiducial' method (green shaded area) reproduces the scatter in the 3 pkpc aperture mass at higher redshifts. The scatter in the 1,5 and 70 pkpc apertures is also reproduced (not shown).

\begin{figure}
\includegraphics[width=\columnwidth]{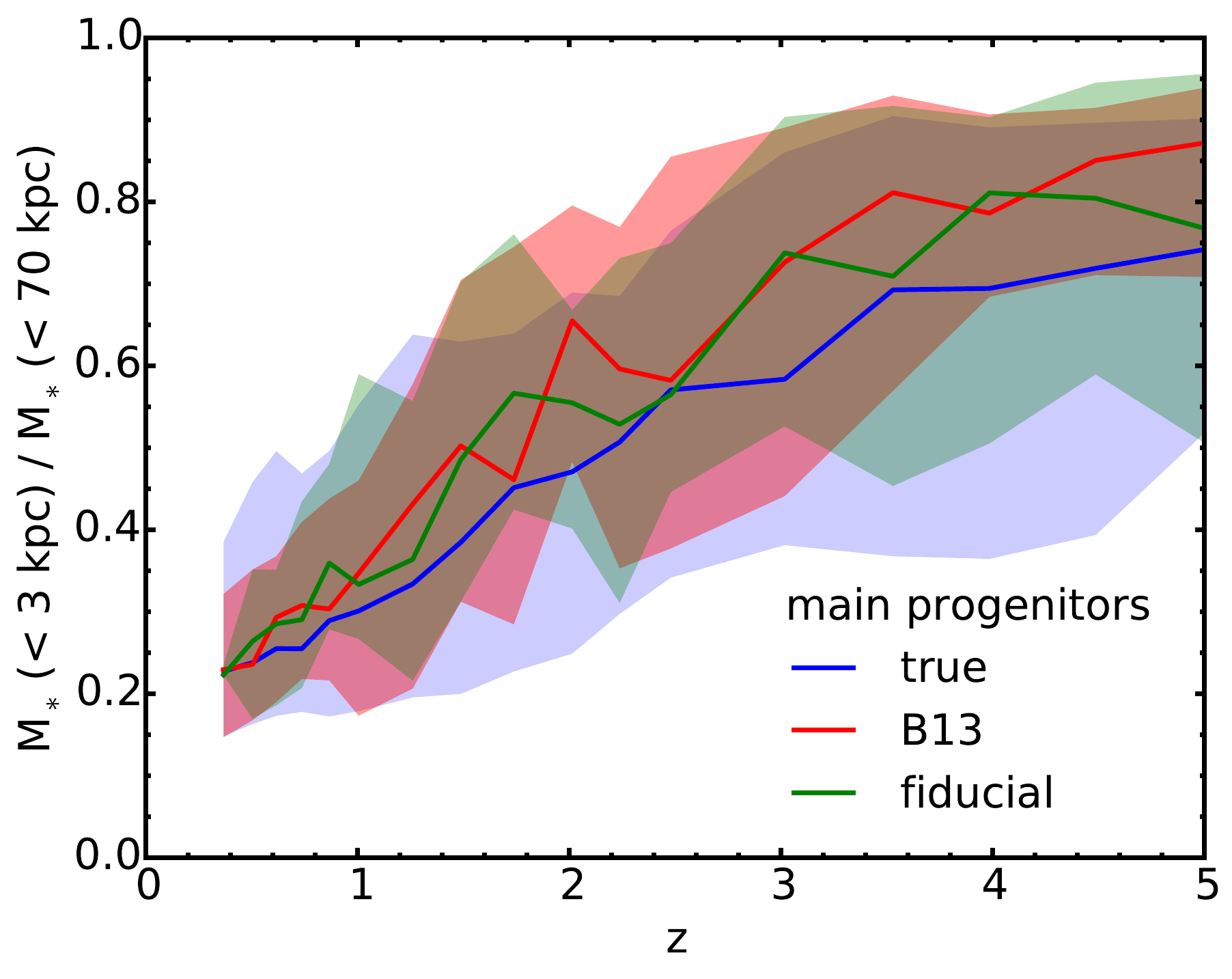}
\caption{The fraction of stellar mass that resides in the central 3 pkpc as a function of redshift. Solid lines denote the median and shaded regions denote the 10$\%$-90$\%$ range for the same galaxy selections as in Figs. \ref{Figure1} and \ref{Figure2}. Both methods reproduce the general trend, but slightly overestimate the fraction at all redshifts. The `fiducial' method reproduces the scatter at high redshift better than the `B13' method.}
\label{Figure3}
\end{figure}

Figure \ref{Figure3} shows the evolution of the median galaxy stellar mass fraction within a 3 pkpc radius (blue curve). This fraction evolves from $\approx 0.74$ at $z=5$ to $\approx 0.22$ at $z= 0.37$, clearly indicative of an inside-out growth history. Similar inside-out behaviour is observed for the other apertures (not shown). The fraction for the 1 pkpc and 5 pkpc apertures evolves from $\approx0.36$ to $\approx0.05$ and from $\approx0.81$ to $\approx0.33$ respectively. The `B13' method (red curve) slightly overestimates this inside-out growth trend for individual galaxies, as does the `fiducial' method (green curve). However the `fiducial' method succeeds better in reproducing the scatter at high redshifts (indicated by the 10$\%$-90$\%$ shaded regions).

The evolution of the average stellar density profile is dominated by the most massive main progenitors at each redshift. This evolution is estimated well by the `B13' method, despite the fact that the galaxies in the `B13' selection are neither representative of the true main progenitors' mass (Fig. \ref{Figure2}) nor of their evolution (Fig. \ref{Figure3}). The `fiducial' method improves on all these points, but might be difficult to implement observationally, because it requires deeper observations (typically an order of magnitude fainter at $z=5$). However, when the objective is to retrieve the average growth of the stellar profile density, we expect the sampling of scatter in mass at the high-mass end to be of much greater importance than that at the low-mass end. Thus, a galaxy selection similar to our `fiducial' model, but with a low mass cut could be a viable alternative.

\section{Comparison with observations}

Up to this point we have used the EAGLE simulation to validate the use of a CND matching technique based on B13 to retrieve the history of the stellar density profile of massive galaxies. In this section we will directly compare EAGLE to observations obtained with this technique.

\begin{figure*}
\includegraphics[height=0.372\textwidth]{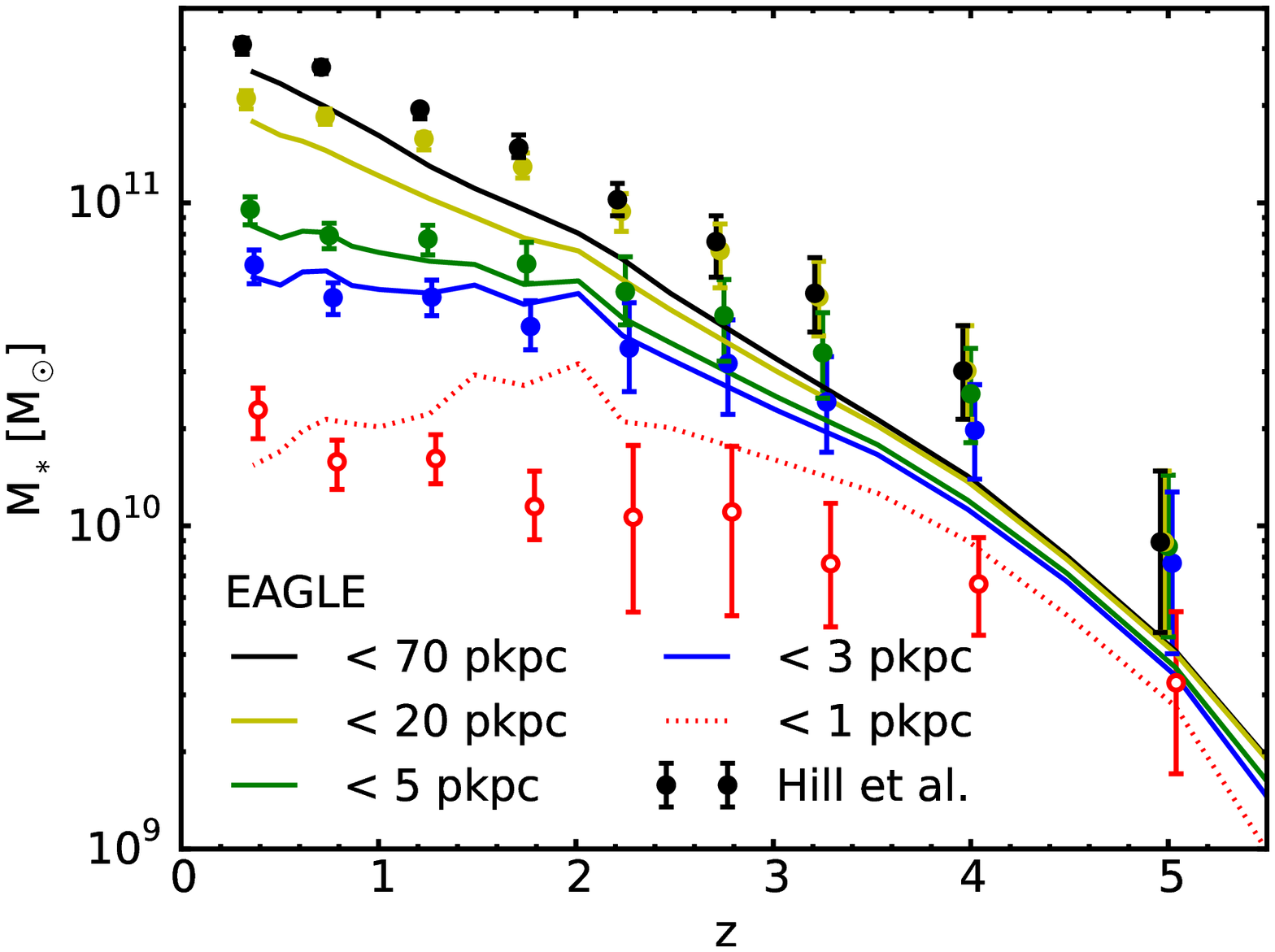}
\includegraphics[height=0.38\textwidth]{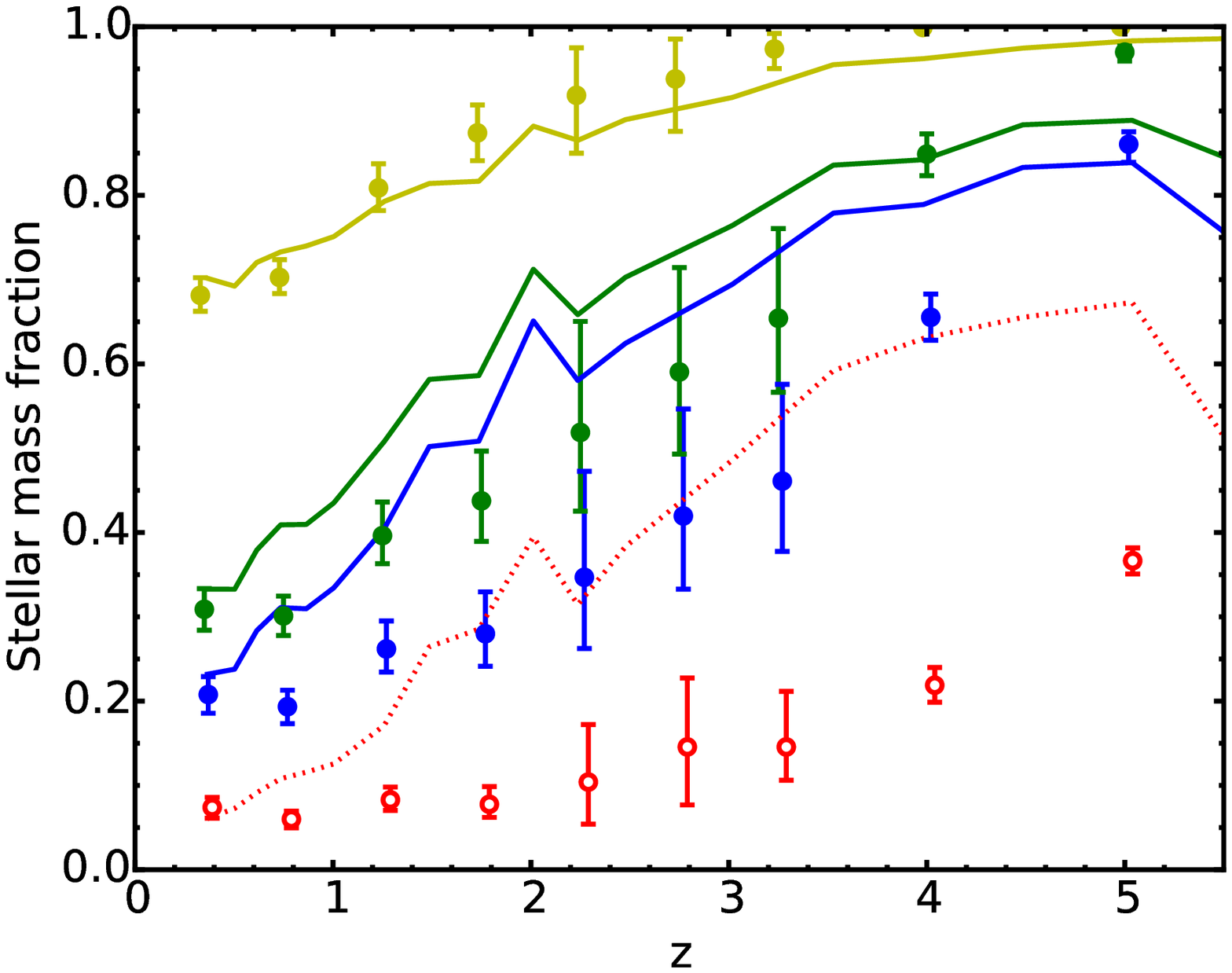}
\caption{A comparison of the evolution of the aperture masses in the EAGLE simulation (curves) to the observations from \protect{\citet{Hill16}} (data points). Galaxies are selected both in the simulation and in the observations at the median CND from B13. This CND corresponds to a total stellar mass of $10^{11.5} {\rm M_{\odot}}$ at $z=0.35$ in the observations (black data points) and to a slightly lower mass in EAGLE (black curve). Different colours denote the stellar mass within different 3D apertures. The red dashed curves and open data points should be treated with caution because they are probing a regime that is unresolved. In the right panel the masses are normalised to the total stellar mass at each redshift in order to show the inside-out growth trend. The error bars in the right panel are from S\'ersic fitting. In the left panel the error bars also include uncertainties in the SED-fitted masses from photometry, uncertainties in photometric redshifts and cosmic variance. Overall there is good agreement between the simulation and observations. Both indicate an inside-out growth history over $0\lesssim z \lesssim 5$. However, at $z\lesssim 2$ the inside-out growth trend in EAGLE is more prominent than observed.}
\label{Figure4}
\end{figure*}

Figure \ref{Figure4} compares the evolution of aperture masses in EAGLE with those obtained by \citet{Hill16}. The observations comprise stacks of UltraVISTA (Data Release 3, not yet public) \citep{McCracken12,Muzzin13b} and CANDELS \citep{Brammer12,Skelton14} images. These stacks are fitted with a S\'ersic profile and then de-projected, in order to compare them to the 3D aperture masses in EAGLE. For details see \citet{Hill16}. We have tried to keep the analysis of the simulation and observations as similar as possible.

Both the simulation and observations are evaluated along the evolving median CND of B13 that corresponds to a stellar mass of $10^{11.5} {\rm M_{\odot}}$ at $z=0.35$. The observational errors take into account the uncertainty in the S\'ersic parameters, in the SED-fitted masses from photometry, in the photometric redshifts, and cosmic variance. Both the SED-fitting and the EAGLE simulation assume a Chabrier IMF. The stellar mass in EAGLE does not include stars that reside in other subhalos (e.g. satellite galaxies). Similarly, in the observations, satellite galaxies and interlopers are masked. The total stellar mass in EAGLE is measured within a 3D aperture of 70 pkpc. In the observations the total stellar mass of the stack is taken to be the sum of the individual catalogue galaxy masses from SED-fitting. A single $M/L$ ratio is assumed for the stack such that the total stellar mass resides within a 75 pkpc 3D deprojected aperture. This might induce an error in cases where the catalogue mass includes stellar light from outside 75 pkpc or in cases where stellar light within 75 pkpc is lost in the noise. $M/L$ gradients, which could contribute at the $\approx 0.2$ dex level, are not taken into account, whereas in EAGLE these are included intrinsically, as the simulation directly traces stellar mass.

Finally, caution should be taken in interpreting the results for the stellar mass within the 1 pkpc aperture (red dashed lines and open data points). In EAGLE this is close to the resolution limit of the simulation. In the observations, for most galaxies the central pkpc is below the resolution limit and hence the mass within this aperture is mostly driven by the S\'ersic fit.

Figure \ref{Figure4} shows overall agreement between observations and simulation. The difference between EAGLE and observations in the total stellar mass at the B13 CNDs (black curve and data points) diverges to approximately a factor of 2 at high redshifts, albeit at $1\sigma$ level at $z=5$. For the 1 pkpc aperture (red dashed curve and open data points) the trend over the entire redshift range is similar, but EAGLE shows a declining 1 pkpc aperture mass for $0\lesssim z\lesssim2$. A comparison with Figure \ref{Figure1} (left panel) shows that this is mainly a feature of the B13 selection and not of the true main progenitors. Both simulation and observations show an inside-out growth history over $0\lesssim z \lesssim 5$, indicated by the declining stellar mass fraction within the smaller apertures towards lower redshifts in the right panel. For $0\lesssim z\lesssim 2$ the observations hint at a less pronounced inside-out growth trend than the EAGLE simulation.

In future an improvement on the comparison of the stellar density profile growth in observations and simulation can be made by taking into account $M/L$ gradients in the observations or by making virtual observations of the simulation and comparing light profiles directly. If the growth is aimed to be representative of main progenitor growth then improvements can be made by also sampling the scatter in stellar mass at high redshifts (our fiducial method).

\section{Conclusions}
\label{SectionConclusions}

We used the EAGLE hydrodynamical simulation to determine whether the growth of the stellar density profile of the main progenitors of $M_{*}=10^{11.5} {\rm M_{\odot}}$ galaxies can be reliably estimated with the cumulative number density (CND) matching technique. Our conclusions are as follows.
\begin{itemize}
\item{The average stellar mass growth within 1, 3 and 5 pkpc 3D apertures is well approximated by selecting galaxies that follow the evolving CND of B13 (Fig. \ref{Figure1}). This suggests that using the `B13' method to account for merging is a reasonable way to study the radially dependent build-up of stellar mass for these massive galaxies. The expected errors in the aperture masses grow to a factor of $\approx 2$ at $z=5$.}
\item{The CND method can be improved by also sampling the expected scatter in the main progenitor stellar masses. We assume a lognormal form for this scatter, fitted to the $68^{\rm th}$-percentile range of B13. This `fiducial' method reduces the error in the retrieved aperture masses by a factor of $\approx 2$, especially at high redshifts, where the expected scatter in main progenitor stellar masses is large (Fig. \ref{Figure1}).}
\item{Although the `B13' method succeeds in reproducing the average stellar mass growth within different apertures, it reproduces neither the scatter in these aperture masses (Fig. \ref{Figure2}) nor the scatter in the central concentration of stellar mass (Fig. \ref{Figure3}). The fact that the average aperture masses are reproduced is somewhat of a coincidence, since it is the average for a selection of galaxies that is neither representative in mass nor in the evolution of the true main progenitor sample.}
\item{Sampling the scatter in stellar mass with the `fiducial' method yields results that are more representative of the true main progenitors' aperture masses (Fig. \ref{Figure2}) and central stellar concentrations (Fig. \ref{Figure3}).}
\end{itemize}

Finally, we compared the evolution of 1, 3, 5 and 20 pkpc 3D mean aperture masses in EAGLE to the stacked UltraVISTA and CANDELS observations from \citet{Hill16} (Fig. \ref{Figure4}). Both in the simulation and in the observations we sample the galaxies along the evolving B13 median CND. Overall the simulation and observations agree remarkably well. Both indicate an inside-out stellar growth history over $0 \lesssim z \lesssim 5$. However, in EAGLE most of the relative inside-out growth happens for $z \lesssim 2.5$, whereas in the observations the inside-out aspect of the growth is more prominent at $z \gtrsim 2.5$.

\section*{Acknowledgements}

This work was supported by the Netherlands Organisation for Scientific Research (NWO), through VICI grant 639.043.409 and by the European Research Council under the European Union's Seventh Framework Programme (FP7/2007- 2013) / ERC Grant agreement 278594-GasAroundGalaxies.

\bibliographystyle{mn2e} 
\bibliography{Bibliography}

\label{lastpage}
\end{document}